# Graphene-assisted preparation of large-scale single crystal Ag(111) nanoparticle arrays

Yibo Dong[a], Yiyang Xie[a,*], Liangchen Hu[a], Chen Xu [a,*], Weiling Guo[a], Guanzhong Pan[a], Qiuhua Wang[a], Fengsong Qian[a], and Jie Sun [a,b,*]

[a] Key Laboratory of Optoelectronics Technology, College of Microelectronics, Beijing University of Technology, Beijing 100124, China

[b] National and Local United Engineering Laboratory of Flat Panel Display Technology, Fuzhou University, Fuzhou 350116, China

*Corresponding author. E-mail: xieyiyang@bjut.edu.cn (Yiyang Xie), xuchen58@bjut.edu.cn (Chen Xu), jie.sun@chalmers.se (Jie Sun).

**ABSTRACT**

Surface plasmon resonance of metal nanostructures has broad application prospects in the fields of photocatalysis, optical sensing, biomarkers and surface-enhanced Raman scattering. This paper reports a graphene-assisted method for preparing large-scale single crystal Ag(111) nanoparticle arrays based on ion implantation technique. By surface periodic treatment and annealing of the implanted sample, regularly arranged Ag nanoparticles can be prepared on the sample surface. A new application for graphene is proposed, that is, as a perfect barrier layer to prevent metal atoms from evaporating or diffusing. All the Ag NPs show (111) crystal orientation. Besides, the Ag atoms are covered by graphene immediately when they precipitate from the substrate, which can prevent them from being oxidized. On the basis of this structure, as one of the applications of metal SPR, we measured the Raman enhancement effect, and found that the G peak of the Raman spectrum of graphene achieved about 20 times enhancement.

# 1. Introduction

Surface plasmon resonance (SPR) refers to the resonance of incident light with the vibration of free electron masses on the surface of metal nanostructures.[1-3] Surface plasmon optics of metal nanostructures has broad application prospects in the fields of photocatalysis, optical sensing, biomarkers, medical imaging, solar cells, and surface-enhanced Raman scattering (SERS).[1] Noble metal nanoparticles (NPs), such as Au, Ag and Pt, have strong local surface plasmon resonance effect, which has attracted wide attention in recent years. They can exhibit strong spectral absorption in the ultraviolet, visible and infrared bands. The peak of the absorption spectrum depends on the microstructural characteristics of the NPs, such as the composition, shape, size and structure.[4] At present, there are many ways to prepare metal NPs, such as electron beam lithography, chemical synthesis, annealing of deposited metal films, nanoimprint lithography and ion implantation followed by annealing.[5-13]

Metal ion implantation followed by annealing is one of the common methods for preparing metal NPs.[10-13] During annealing, the metal atoms implanted into the substrate will aggregate to form metal NPs inside the substrate. This method can obtain dense metal NPs in a large area. At the same time, the lattice of the metal NPs is better due to the high temperature annealing during the preparation process. However, the disadvantages of this method are also obvious. First, the irregular size and distribution of metal NPs make the controllability of this method very poor. Second, metal atoms will precipitate from the substrate and evaporate under the action of high temperature, resulting in the loss of metal atoms. Third, because of the fear of evaporation of Ag, the annealing temperature can not be too high, so the size of Ag NPs is small (tens of nanometers).

Graphene, as the first two-dimensional material, has attracted worldwide attention for its many excellent properties since its discovery in 2004.[16] As one of its properties, however, the application of ultrahigh mechanical strength (~1 TPa) of graphene is rarely reported.[15]

In this work, we have proposed a graphene-assisted method for preparing large-scale Ag NPs based on ion implantation technique. $Ag^+$ implantation followed by annealing is used to prepare Ag NPs on the $SiO_2$/Si substrate. For the first time, graphene is used as a perfect barrier layer to prevent metal atoms from evaporating, so the annealing temperature can be higher than the melting point of Ag. Besides, by surface periodic treatment (SPT) of the substrate, the obtained Ag NPs can be regular arranged and form Ag nanoparticle arrays (Ag NAs). The obtained Ag NPs are analyzed by optical, scanning electron microscope (SEM), atomic force microscope (AFM) and X-ray diffraction (XRD) measurements. This method basically solves all the problems mentioned above of metal NPs prepared by traditional ion implantation technique, and realizes efficient preparation of Ag NAs with large area, regular arrangement and uniform size. The obtained Ag NPs are all single crystal and show (111) crystal orientation. Ag oxidation has always been a major obstacle to the application of Ag NPs in SPR. This method solves the problem of long-term storage of Ag NPs. When the Ag NPs are prepared, they are wrapped with graphene to block the air and prevent Ag from being oxidized. Not only Ag, this method can achieve the preparation of a variety of metal NPs. On the basis of this structure, as one of the applications of SPR, we measured the Raman enhancement effect. The G peak of the Raman spectrum of graphene achieved about 20 times enhancement.

## 2. Experimental Section

### 2.1 Preparation of Ag NPs

First, Ag atoms are implanted into the shallow surface of the substrate by ion implantation (Figure 1a). The implanted energy is 30 keV corresponding to the implanted depth of about 22 nm (Figure S1, Supporting information), and the implanted dose is $5\times10^{15}$/cm$^2$. The substrate we used are heavily doped n-type Si with 300 nm SiO$_2$. Then, a monolayer graphene is transferred onto the substrate (Figure 1b). Graphene transfer we used is a common wet transfer process. Annealing at 1050 °C for 15 min in a H$_2$/Ar (40 sccm/960 sccm) atmosphere (Figure 1c). Finally, we will obtain Ag NPs on the substrate surface (Figure 1d). As shown in Figure 1e and f, the arrangement of these Ag NPs is disordered.

**2.2 SPT**

First, SiO$_2$ nanospheres are transferred onto the substrate (Figure 2a). The transfer step of the SiO$_2$ nanospheres is as follows: 1. Preparing a beaker with deionized water. 2. A solution mixed with SiO$_2$ nanospheres was dropped into the sidewall of the beaker to allow the solution to slowly flow into the deionized water. Due to the surface tension of water, the SiO$_2$ nanospheres will float on the surface of the water. 3. The substrate is first immersed in water using a pulling machine, and then pulled up. The SiO$_2$ nanospheres will be attached to the surface of the substrate. After transfer, the SiO$_2$ nanospheres are used as masks to dry etch the substrate in order to separate the implanted region into several regularly arranged small regions (Figure 2b). The gas used for etching is CHF$_3$ (100 sccm), and the etching power is 300W. The etching time is 50 s, and the corresponding etching depth is about 40 nm. Then, removal of SiO$_2$ nanospheres by ultrasound in deionized water.

After SPT, we transfer graphene onto the substrate and carry out an annealing (Figure 2c and d). Finally, Ag NAs with regular arrangement the same as the SiO$_2$ nanospheres can be obtained.

**2.3 Material Characterization**

The surface morphology of Ag NAs was characterized by optical observation (Olympus BM51 optical microscope), SEM (Zeiss Merlin scanning electron microscope operating at 15 kV) and AFM (Bruker scanning probe microscope, Dimension FastScan). XRD result is obtained using a Shimadzu X-ray diffractometer (XRD-7000). Raman spectra were obtained using a Horiba LabRAM HR Evolution Raman microscope (532 nm laser).

**3. Results and discussion**

**3.1 Preparation of large-scale Ag NPs**

Annealing is an important part of this technique. During annealing, Ag atoms will diffuse in the substrate and precipitate from the substrate when the temperature is cooling down. Graphene acts as a perfect barrier to prevent the evaporation of Ag atoms. Since no substance can pass through graphene (except $H^+$),[14] when Ag atom precipitates from the substrate, it will not evaporate directly, but will be bound by graphene at the interface between graphene and substrate and agglomerate to form Ag NPs. Meanwhile, because of its strong mechanical strength, graphene can not be torn by Ag vapor even if it is only a single atomic layer thickness. In addition, the stretchability and flexibility of graphene also play an important role. Therefore, the annealing temperature (~1050 °C) can be much higher than the melting point of nano-silver. Without graphene, the Ag atoms will evaporate completely and no Ag NPs are observed (Figure S2, Supporting information). Figure 1e is an optical image of the final Ag NPs at 1000 magnification. The white spots in the image are the Ag NPs. Figure 1f is a SEM image of Ag NPs. Figure 1g is a SEM image of a single Ag NP with a diameter of about 150 nm. We can clearly see that graphene is supported by the Ag NP, thus graphene folds are forming around the Ag NPs. Not limited to Ag, we also prepared Cu NPs using the same method (Figure S3, Supporting information).

Compared with traditional ion implantation and annealing method, with graphene barrier, we can easily prepare large-area Ag NPs on the substrate surface. Besides, because of the high annealing temperature, the size of obtained Ag NPs can be much larger than previous reports.[12,13] The graphene can provide a good protection for Ag NPs from oxygen and ensure their long-term stability.

It should be pointed out that the graphene we used is commercial polycrystalline graphene grown on Cu foil. Meanwhile, the transfer process may lead to graphene breakage. Therefore, at the grain boundaries or broken sites, graphene can not effectively block the evaporation of Ag atoms, leading to no Ag NPs formed at some area covered by graphene. This problem can be solved by some methods. We found that multiple transfer of monolayer graphene can significantly reduce the area without Ag NPs (Figure S4, Supporting information). At the same time, there have been many reports about the growth of large-area single-crystal graphene.[17,18] Therefore, we believe the use of single crystal graphene may solve this problem.

**3.2 Regular arrangement of Ag NPs**

One of the most important problems in the preparation of metal NPs by annealing is the failure to form regular arrangement. However, in this work, the Ag NPs can be periodically aligned by SPT. As shown in Figure 2a-d, we transferred $SiO_2$ nanospheres on the surface of the sample. The diameter of $SiO_2$ nanospheres is about 600 nm. Due to the influence of water surface tension in the transfer process, the transferred $SiO_2$ nanospheres are regularly arranged on the substrate, as shown in Figure 1e. After that, we used $SiO_2$ nanospheres as masks to dry etch the sample, and divided the surface of the sample into many cylindrical mesas, as shown in Figure 1f. The etching depth is about 40 nm, which is deeper than the implanted depth, so the implanted region is completely separated into many 600-nm diameter circular regions, each of which is independent of each other. After SPT, we transferred graphene and annealed the sample. Interestingly, the obtained Ag NPs show a regular alignment as same as the $SiO_2$ nanospheres. As shown in Figure 2g, the left image shows the Ag NPs with SPT, while the right image shows the Ag NPs without SPT. This result excites us because this method can realize the preparation of large-area Ag NPs in mechanism. Compared with electron beam lithography, this method has advantages in efficiency and cost. Besides, we can not only use $SiO_2$ nanospheres as masks, but also use nanoimprint and other means to realize SPT with different arrangements.

Figure 3a and b are the SEM images of the Ag NAs. It can be seen that the arrangement of Ag NPs is the same as that of $SiO_2$ nanospheres in Figure 2e. They are all regular hexagonal. The distance between two adjacent Ag NPs is about 600 nm (white dotted line in Figure 3a), which is consistent with the diameter of $SiO_2$ nanospheres, proving that the regular arrangement of Ag NPs is caused by the periodic cylindrical mesas on the substrate surface. In order to obtain the information about the relative positions of Ag NPs and cylindrical mesas, we removed the graphene by oxygen plasma. As shown in Figure 3c, the circle marked by the black dotted line is the position of the cylindrical mesa. All the Ag NPs are located at the center of the cylindrical mesas, which means that the Ag atoms in each cylindrical mesa form a single Ag NP. Figure 3d is the three-dimensional sketch of the Ag NAs. Figure 3e and f are the AFM images of Ag NAs before (e) and after (f) removal of graphene. It should be noted that Ag NPs will also be etched by oxygen plasma, resulting in changes in surface morphology.

At the same time, we found that the size of Ag NPs prepared with SPT is more consistent. We have counted the size distribution of Ag NPs in the same size area shown in Figure S5a and b (Supporting information). Since Ag NPs are not standard circular, we count the width of each Ag NP, i.e. the projection length in the x-axis direction. As shown in Figure 4a, without SPT, the size distribution is very wide and the average size is relatively small. On the contrary, the size distribution of Ag NPs is concentrated around 144 nm with SPT, and the average size of Ag NPs is relatively larger. This is because after SPT, the diameter of each cylindrical mesa is basically the same, so the implanted dose of Ag atoms in each cylindrical mesa is almost the same. Therefore, the diameter of Ag NPs formed on each cylindrical mesa is almost the same. This phenomenon also gives us an inspiration that we can control the size of the Ag NPs by controlling the implanted dose and the diameter of the cylindrical mesa.

Figure 4b shows the XRD results. We can see that there is only one diffraction peak corresponding to (111) orientation of Ag,[23] which proves that almost all the NPs are (111) orientation. Meanwhile, our annealing temperature (1050 °C) is much higher than the melting point (<960 °C) of nano-silver. It is almost impossible for the obtained Ag NPs to have multiple grains. From the SEM image (Figure 1g), we also did not observe grain boundaries. Therefore, we can draw a conclusion that all the Ag NPs are Ag (111) single crystals.

**3.3 SERS of the Ag NAs**

SERS is an important application of metal SPR. SERS technology overcomes the weakness of traditional Raman signal and can increase the Raman intensity by several orders of magnitude. In order to get a strong enhancement signal, first of all, we need a good substrate. Graphene, as a two-dimensional ultra-thin carbon material, is easy to adsorb molecules and can be said to be a natural substrate. When some molecules are adsorbed on the surface of graphene, the Raman signal of the molecules will be significantly enhanced. Recently, there have been many reports on graphene SERS.[8,19-22] Metal plasma nanostructures can enhance Raman signals by generating strong local electric fields. Therefore, the combination of graphene and noble metal plasma nanostructures can effectively enhance Raman signal. Our technique just combines graphene and Ag NAs. Thus, as one of the applications, we measured the Raman enhancement effect of the Ag NAs.

Figure 5a shows the Raman enhancement results of the Ag NAs. Raman spectra of three kinds of graphene were measured respectively. First, we measured the Raman spectrum of graphene before annealing (black curve in figure 5a, Gr before annealing). Then, there are no Ag NPs formed in some graphene-covered areas after annealing due to the defects or transfer-caused damages of graphene. We also measured the Raman signal in these areas (blue curve in figure 5a, Gr after annealing). Finally, we measured the Raman signal of graphene with Ag NAs (red curve in figure 5a, Gr/Ag after annealing). As we can see, the G peak of graphene with Ag NAs increased by about 20 times compared with that of graphene before annealing. This enhancement effect is obvious. Meanwhile, we have also noticed two obvious phenomena. First, the Raman signal of graphene is still enhanced in the region without Ag NAs after annealing. This is mainly due to the fact that Ag atoms do not precipitate completely. Some Ag atoms aggregate inside the substrate to form NPs.[12] Therefore, there are still some Raman enhancement effects in these regions. Second, we found that the D peak of graphene increased significantly after annealing, representing the increase of graphene defects, which also led to the decrease of the 2D peak. The increase of graphene defects may be due to the deformation of graphene caused by high temperature annealing and precipitation of Ag atoms. For SERS, graphene is mostly used just as a substrate, so we think that the increase in graphene defects is tolerable. We placed the sample in the air for 10 weeks and measured it again. It was found that the Raman signal of graphene can still be enhanced. By optical observation, the color of the Ag NPs did not change and remained silvery white (Figure S6, Supporting information). Combined with the results, we can conclude that the Ag NPs prepared by this method can be stored for a long time, and graphene protects the Ag NPs from oxidation. Figure 5d shows the G-peak Raman mapping of the graphene in the area shown in figure 5c. Figure 5c is a dark-field optical image of the Ag NAs. It can be seen that the Raman enhanced region fits well with the region of the Ag NAs. In figure 5c, there is a strip region without Ag NAs, and the Raman signal in this region also shows obvious weakness.

## 4. Conclusion

We propose a graphene-assisted method to fabricate Ag NAs. Ag NPs were prepared on the substrate surface by $Ag^+$ implantation and annealing. Graphene acts as a barrier to prevent the evaporation of Ag atoms. In terms of mechanism, this method can realize the preparation of wafer-scale Ag NPs. After SPT with the substrate, a periodic arrangement of Ag NAs can be obtained. Theoretically, the diameter of the obtained Ag NPs can be controlled by the implanted dose and the diameter of the $SiO_2$ nanospheres. The Ag atoms are covered by graphene immediately when they precipitate from the substrate, which can prevent them from being oxidized. As one of the applications of metal SPR, the Raman enhancement effect of the structure was measured, and the G peak of graphene was enhanced about 20 times. After long-term storage, Ag NPs are still not oxidized, which proves that the graphene has played a perfect protective role. This preparation method of Ag NAs has potential applications in SERS, optical sensing, photocatalysis and biomarkers in the future. At present, there is still some room for improvement in this work. The defects of graphene and the damage caused by wet transfer result in the absence of Ag NPs in some areas. In the future, we will try to use large-area single-crystal graphene as the barrier layer and improve the graphene transfer process to ensure the integrity of the graphene.

**Acknowledgements**

We acknowledge support from the National Key R&D Program of China (2018YFA0209000), the National Natural Science Foundation of China (11674016 and 61874145), the Beijing Natural Science Foundation (4172009), the Beijing Municipal Commission of Science and Technology (Z161100002116032), the Beijing Municipal Commission of Education (KM201810005029).

[23] Ariosa, D.; Cancellieri, C.; Araullo-Peters, V.; Chiodi, M.; Klyatskina, E.; Janczak-Rusch, J.; Jeurgens, L. P. H. Modeling of Interface and Internal Disorder Applied to XRD Analysis of Ag-Based Nano-Multilayers. *ACS Appl. Mater. Interfaces* **2018**, 10, 20938−20949.15

**Figures:**

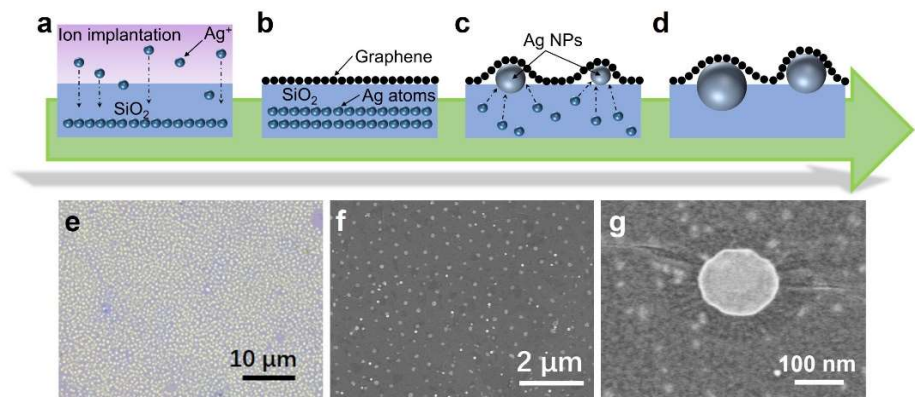

**Figure 1.** (a-d) Sketch of graphene-assisted preparation of large-area Ag NPs. (a) Ag$^+$ implantation. (b) Transferring monolayer graphene onto the substrate. (c) Annealing at 1050 °C for 15 min. During annealing, Ag atoms will diffuse to the surface of the substrate and aggregate to form Ag NPs. (d) Finally, Ag NPs were obtained on the surface of the substrate. (e) Optical image of the Ag NPs. (f) SEM image of the Ag NPs. (g) SEM image of a single Ag NP. Graphene folds can be clearly seen around the Ag NP.

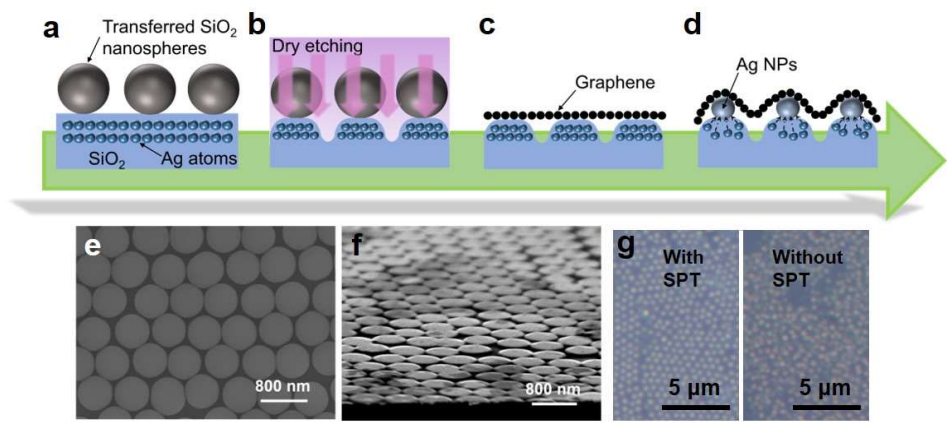

**Figure 2.** (a) Transferring SiO$_2$ NPs. (b) Dry etching of the substrate. (c) Transferring monolayer graphene. (d) Annealing at 1050 °C for 15 min to obtain regularly arranged Ag NAs. (e) SEM image of SiO$_2$ NPs after transferring onto the substrate. (f) SEM image of the cylindrical mesa on the substrate. This image was observed at an angle of 75 degrees. (g) Optical images of Ag NPs prepared on the substrates with and without SPT respectively.

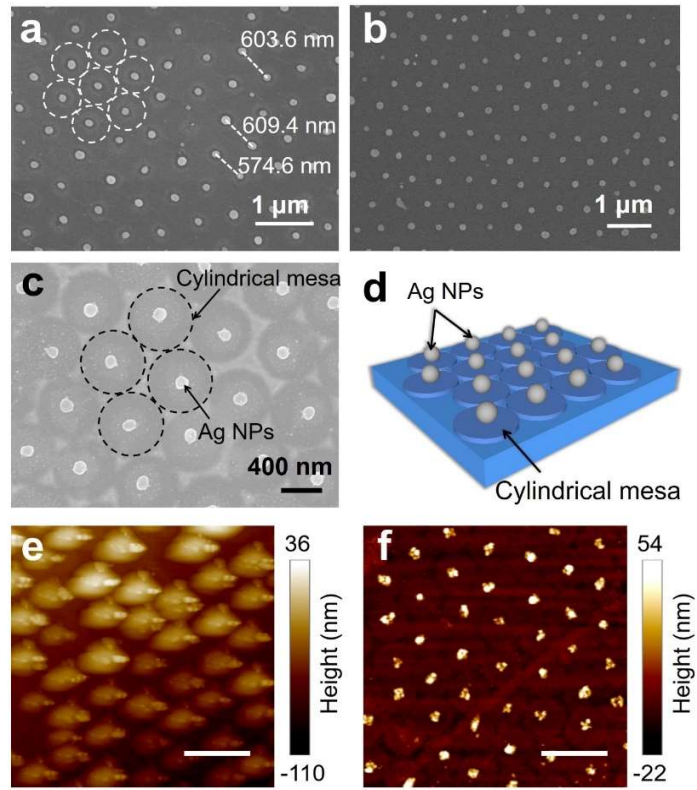

**Figure 3.** (a)(b) SEM images of the Ag NAs obtained by annealing after SPT. The distance between adjacent Ag NPs is about 600 nm. (c) SEM image of the Ag NAs after etching the graphene. The circle marked by the black dotted line in the image is the position of the cylindrical mesa. (d) Three-dimensional sketch of the Ag NAs. (e)(f) AFM images of the Ag NAs before (e) and after (f) etching the graphene. The scale bars in figure 3e and f stand for 1 μm.

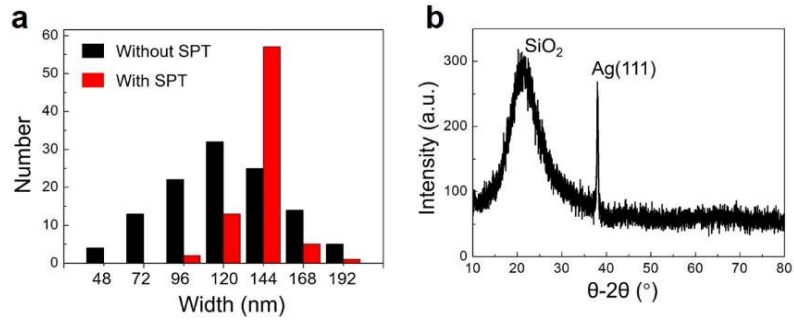

**Figure 4.** (a) Width distribution of Ag NPs. Each abscissa represents the range of positive and negative 12 nm of the number. For example, number 144 represents a range of 132 nm to 156 nm. (b) XRD patterns of the obtained Ag NAs.

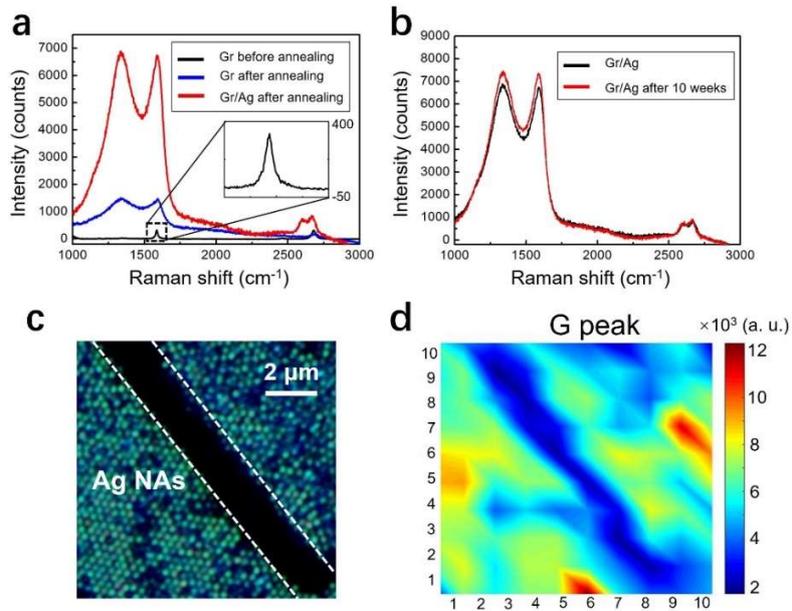

**Figure 5.** (a) Raman spectra of the graphene. The black curve represents the graphene Raman signal before annealing. The blue curve represents the graphene Raman signal after annealing and without Ag NAs. The red curve represents the graphene Raman signal after annealing and with Ag NAs. (b) The Raman signal of the graphene with Ag NAs before and after 10 weeks storage in air. (c) Dark-field optical image of the Ag NAs. (d) The G-peak Raman mapping of the graphene in the region shown in (c).